# Unveiling User Perceptions in the Generative AI Era: A Sentiment-Driven Evaluation of AI Educational Apps' Role in Digital Transformation of e-Teaching

Adeleh Mazaherian
Educational Sciences & Psychology Department
Islamic Azad University
Isfahan, Iran
adeleh.mazaherian@iau.ir

Erfan Nourbakhsh
Artificial Intelligence Department
University of Isfahan
Isfahan, Iran
erfannourbakhsh2001@gmail.com

*Abstract—* The rapid integration of generative artificial intelligence into education has driven digital transformation in e-teaching, yet user perceptions of AI educational apps remain underexplored. This study performs a sentiment-driven evaluation of user reviews from top AI ed-apps on the Google Play Store to assess efficacy, challenges, and pedagogical implications. Our pipeline involved scraping app data and reviews, RoBERTa for binary sentiment classification, GPT-4o for key point extraction, and GPT-5 for synthesizing top positive/negative themes. Apps were categorized into seven types (e.g., homework helpers, math solvers, language tools), with overlaps reflecting multifunctional designs. Results indicate predominantly positive sentiments, with homework apps like Edu AI (95.9% positive) and Answer.AI (92.7%) leading in accuracy, speed, and personalization, while language/LMS apps (e.g., Teacher AI at 21.8% positive) lag due to instability and limited features. Positives emphasize efficiency in brainstorming, problem-solving, and engagement; negatives center on paywalls, inaccuracies, ads, and glitches. Trends show that homework helpers outperform specialized tools, highlighting AI's democratizing potential amid risks of dependency and inequity. The discussion proposes future ecosystems with hybrid AI-human models, VR/AR for immersive learning, and a roadmap for developers (adaptive personalization) and policymakers (monetization regulation for inclusivity). This underscores generative AI's role in advancing e-teaching by enabling ethical refinements that foster equitable, innovative environments. The full dataset is available here(https://github.com/erfan-nourbakhsh/GenAI-EdSent).

*Index Terms--* Generative AI, e-Teaching, sentiment analysis, educational apps, digital transformation

## I. Introduction

The Generative artificial intelligence (GenAI) has reshaped education, advancing digital transformation in e-teaching through enhanced personalization, accessibility, and efficiency [1]. GenAI supports dynamic features such as automated tutoring and adaptive assessments, as seen in the rise of AI-integrated educational apps (ed-apps) on platforms such as the Google Play Store[1], addressing needs ranging from homework assistance to language learning [4]. Yet, a key gap remains in grasping user perceptions in real-world use. This study fills this gap through sentiment analysis of user reviews, uncovering strengths, challenges, and implications for equitable e-teaching. Recent surveys reveal varying AI adoption in education, highlighting opportunities and barriers. Students prioritize efficiencies like brainstorming, summarizing, and feedback [2]. Educators focus on strategic uses such as lesson ideas, plans, and simplifying topics [2]. This divergence reflects students' productivity needs versus teachers' pedagogical emphasis. However, most K-12 teachers remain non-users due to concerns like integrity, training, and hurdles [3], [5], [6]. Despite challenges, a dedicated minority drives momentum for personalized learning, underscoring the need for empirical insights to bridge gaps [7], [8].

User perceptions are crucial, revealing satisfaction, ethical, usability, and inclusivity issues in AI's digital transformation role [9], [10]. Traditional evaluations via surveys or interviews are limited by sample size and subjectivity [11]. Instead, app store reviews provide scalable, real-time insights into authentic experiences, including sentiments on accuracy, monetization, and integration [12]. This study uses NLP techniques—RoBERTa for sentiment classification and GPT-4o/GPT-5 for theme extraction—to analyze reviews from 21 top AI ed-apps [13]. Apps are categorized into seven types: AI Quiz & Question Generators, All-in-One Study Companions, Homework Helpers, Math-Focused Solvers, Document/Content Tools, Learning Management Systems (LMS), and Language Learning Apps, highlighting multifunctional overlaps and feedback trends.

The primary objectives are threefold: (1) quantify sentiment distributions and distill key positive/negative themes across app categories; (2) compare performance trends, showing why homework helpers receive high praise while LMS and language apps face criticism; and (3) discuss implications for future AI educational ecosystems, proposing hybrid models integrating AI strengths with human oversight.

---
1. https://play.google.com/

This analysis contributes data-driven recommendations for developers (e.g., enhancing adaptive personalization) and policymakers (e.g., regulating monetization for inclusivity), while addressing untapped potentials like VR/AR for immersive learning [1], [4]. Ultimately, it underscores GenAI's dual-edged impact: democratizing education via accessible tools but risking over-reliance, inequities, and ethical dilemmas if not refined based on user insights [6], [11].

The paper is structured as follows: Section 2 reviews related work, Section 3 details the methodology, Section 4 presents the results, Section 5 discusses the findings, including future implications, and Section 6 concludes.

## II. RELATED WORK

GenAI integration in education has drawn attention for its applications, adoption, and teaching implications. Mittal et al. [1] review its role in e-teaching, including content automation, personalized learning, and admin support; categories include adaptive tutoring, assessment automation, and collaborative platforms; benefits: enhanced accessibility; risks: bias and reduced critical thinking. Alfarwan [4] systematically reviews GenAI in K-12, highlighting use cases like automated feedback and simulations, but notes limited long-term efficacy evidence due to early adoption.

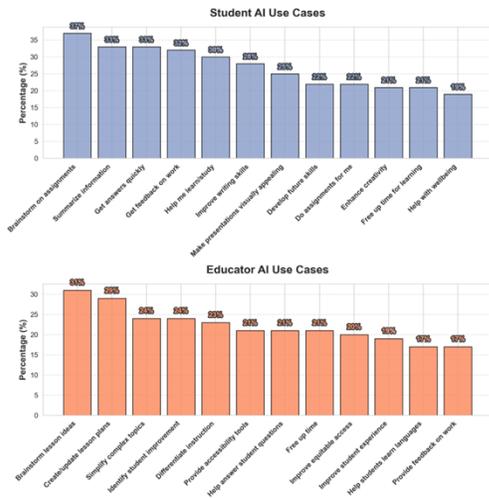

Figure 1. Percentages of primary AI use cases among students and educators for different tasks

Adoption trends reveal a dichotomy between students and educators, per the Microsoft Education report [2]. Students prioritize task efficiencies: brainstorming (37%), summarizing/quick answers (33% each), feedback (32%), study help (30%), writing improvement (26%), presentation visuals (25%), creativity/time management (24% each), skill development/assignment completion (22% each), wellbeing (19%). Educators focus on strategic uses: lesson ideas (31%), lesson plans (29%), simplifying topics/student improvements (24% each), differentiating instruction (23%), accessibility/answering questions/time-saving (21% each), equitable access (20%), student experiences (19%), language/feedback (17% each). Students seek productivity gains, risking over-reliance or plagiarism; educators emphasize pedagogy for diverse needs, requiring training to enhance integration.

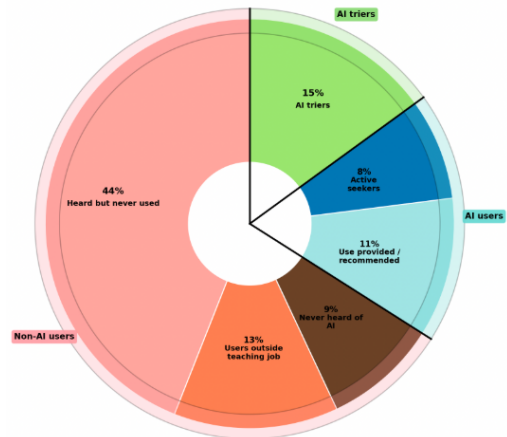

Figure 2. Distribution of teachers' engagement with AI tools in teaching

Figure 2 from a RAND survey [3] reveals teachers' AI engagement: 66% non-users (44% aware but unused in teaching, 13% outside use, 9% unaware). Active users (34%) comprise 15% sporadic triers, 8% proactive seekers, and 11% regular users/recommenders [3]. Caution arises from barriers like training gaps, integrity issues, and constraints, though the minority advances personalized learning and efficiency. Ellis et al. [5] qualitatively examine user experiences, noting GenAI's mix of helpful lesson design support and accuracy issues, and emphasize the need for professional development.

Further studies explore GenAI's applications and user perceptions in higher education. Gabay et al. [6] review GenAI for academic writing, noting benefits like idea generation and editing, but challenges including ethics and misinformation. Vrågård et al. [7] survey educators on GenAI's effects on integrity and learning, highlighting cheating concerns and assessment redesign opportunities. Hu et al. [8] propose a "Socratic Playground" framework for inquiry-based learning via interactive platforms, though user feedback is limited. Haroud and Saqri [9] assess teacher/student views, finding GenAI supports (not replaces) roles, with digital literacy key. Sousa and Cardoso [10] examine student use, showing high adoption for summarization but motivations favoring convenience over deep learning. Mazaheriyan and Nourbakhsh [11] evaluate ethical issues, revealing time-saving drives bounded by authenticity and equity concerns.

Despite these advancements, gaps persist in sentiment-driven evaluations of AI educational apps, particularly through large-scale user review analysis. While prior works rely on surveys or interviews [5], [7], [9], [10], which offer depth but limited scalability, app store reviews provide authentic, voluminous data on real-world perceptions [12]. TweetNLP [12] demonstrates robust NLP for social media sentiment, adaptable to reviews, and GPT-4o [13] enables advanced extraction, yet few studies apply these to ed-apps. This paper addresses these lacunae by conducting a sentiment analysis of Google Play Store reviews for top AI ed-apps, categorizing them functionally and distilling themes to inform future ecosystems, extending beyond descriptive adoption [1]–[4] to prescriptive implications for digital transformation.

## III. METHODOLOGY

To explore user perceptions of AI educational apps in the generative AI era, this study employed a systematic, sentiment-driven evaluation pipeline using NLP and large language models. The method analyzed user reviews from the Google Play Store to gather authentic, real-time feedback on top-rated AI-integrated education tools. It automates sentiment extraction and synthesis, minimizing manual biases while scaling to large datasets. Figure 3 outlines the multi-stage workflow from data collection to theme synthesis.

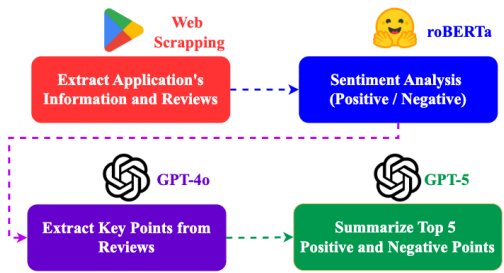

Figure 3. Methodological approach for sentiment-driven evaluation

### A. App Selection and Categorization

We selected 22 prominent AI education apps based on high ratings, most downloads, and explicit generative AI integration, identified via app descriptions and user feedback. The apps are listed below in bullet points with brief overviews:

- **AI Questions Generator:** An AI tool for generating diverse questions from text, including multiple-choice and open-ended, for quizzes and exams.
- **Academi AI: Study & Exam Tutor:** All-in-one AI study companion for homework, quizzes, flashcards, summaries, and resume building.
- **AI Quiz Generator:** An AI app that creates quizzes from text or topics in various formats for teachers and learners.
- **Answer.AI - Your AI tutor:** 24/7 AI tutor for homework across subjects, with explanations, college prep, and premium features.
- **Blackboard:** Mobile LMS app for course access, notifications, assignments, and progress tracking.
- **Brainly:** AI Homework Helper: AI app for instant answers in math and sciences via photo scans, with community and tutor support.
- **Course Hero: AI Homework Help:** AI homework aid with explanations, videos, math solver, and tutor access for various subjects.
- **Edu AI - AI Homework Helper:** Simple AI helper for quick, precise solutions across all homework subjects.
- **Gauth: AI Study Companion:** AI solver for math and sciences with step-by-step help and photo recognition.
- **Help AI: Your Homework With AI:** Versatile AI assistant for homework and queries via text, scan, or voice, with sharing and translation.
- **Homework AI - Math & Essay App:** AI app for math solutions and essay writing, supporting sciences via photo scans.
- **Kahoot! Play & Create Quizzes:** Quiz app for creating and playing engaging games with progress tracking.
- **Mathos AI: Math Helper & Tutor:** AI math solver with explanations, calculators, and support for physics and chemistry.
- **Nerd AI - Tutor & Math Helper:** AI tutor for scanning math, writing, coding, and summaries across subjects.
- **Question.AI - Chatbot&Math AI:** AI chatbot for queries, translations, writing, and study support in multiple languages.
- **Quiz AI: AI Homework Helper:** Learning app for scanning problems, explanations, and organizing digital study materials.
- **Quizard AI - Homework Helper:** AI app for instant answers to quiz questions and math via photo, with explanations.
- **School Hack:** AI app for document chats, summaries, plagiarism checks, and paraphrasing for academic work.
- **Studocu: AI Homework Helper:** Study app with AI solutions, quizzes from docs, and access to notes/exams.
- **StudyX: AI Homework Helper:** AI solver for questions via text or photo, with tutoring and an exam bank.
- **Teacher AI - Language Practice:** AI language teacher for conversations, grammar corrections, and personalized practice.
- **Tutor AI: Learning Assistants:** App for creating custom AI tutors for any subject, with chat and homework help.

Apps were categorized into seven functional types as shown in Table 1 (with overlaps highlighted in red for multifunctional designs). This categorization reflects the evolving AI landscape, emphasizing demand for homework and math aids while identifying growth potential in LMS and language tools.

### B. Data Collection

Data was collected via automated web scraping from the Google Play Store using Python-based tools to extract app metadata (e.g., descriptions, overall ratings) and verbatim user reviews. All reviews up to November 2025 were considered, amassing tens of thousands to millions per app for a robust corpus. This step prioritized popular apps to mirror real-world adoption, ensuring the dataset captured diverse user experiences across educational contexts [12].

### C. Sentiment Analysis and Theme Extraction

The pipeline proceeded in four interconnected stages, as shown in Figure 3:

1) **Binary Sentiment Classification:** Reviews were processed using the RoBERTa model [12], a transformer-based NLP framework pretrained on extensive social media data for robust sentiment detection. It accurately labeled reviews as positive or negative, allowing clear measurement of user satisfaction while still capturing nuances like sarcasm or mixed sentiments.

2) **Key Point Extraction:** Sentiment-labeled reviews were fed into GPT-4o [13], an advanced LLM excelling in zero-shot capabilities, to distill recurring themes, praised features, and pain points. GPT-4o extracted structured insights without fine-tuning, focusing on frequency and contextual relevance.

3) **Theme Synthesis:** Extracted reviews were synthesized using GPT-5, which generated concise summaries of the top 5 positive and negative points per app. Prioritization was based on frequency across reviews and alignment with educational goals, producing actionable overviews.

4) **Aggregation and Trend Analysis:** Sentiments were aggregated to compute positive/negative percentages per app and category, facilitating cross-app comparisons.

| Category | Applications |
|---|---|
| AI Quiz & Question Generators | AI Questions Generator |
| | AI Quiz Generator |
| | Kahoot! Play & Create Quizzes |
| | Quiz AI: AI Homework Helper |
| | Studocu: AI Homework Helper |
| All-in-One Study Companions (Multi-Tool Apps) | Academi AI: Study & Exam Tutor |
| | Answer.AI - Your AI tutor |
| | Nerd AI - Tutor & Math Helper |
| | StudyX: AI Homework Helper |
| | Tutor AI: Learning Assistants |
| | Question.AI - Chatbot & Math AI |
| | Help AI: Your Homework With AI |
| Homework Helpers | Edu AI - AI Homework Helper |
| | Quizard AI - Homework Helper |
| | Course Hero: AI Homework Help |
| | Studocu: AI Homework Helper |
| | Homework AI - Math & Essay App |
| | Gauth: AI Study Companion |
| | Brainly: AI Homework Helper |
| Math-Focused Solvers & Specialized Tools | Mathos AI: Math Helper & Tutor |
| | Gauth: AI Study Companion |
| | Brainly: AI Homework Helper |
| | Homework AI - Math & Essay App |
| Document/Content Tools | School Hack |
| | StudyX: AI Homework Helper |
| | Nerd AI - Tutor & Math Helper |
| Learning Management Systems (LMS) | Blackboard |
| Language Learning Apps | Teacher AI - Language Practice |

Table 1. Categorization of AI educational applications by type (Applications highlighted in red appear in multiple categories due to overlapping functionalities.)

## IV. RESULTS

The sentiment-driven evaluation of user reviews from 20 top AI educational apps on the Google Play Store yielded insights into user perceptions, revealing patterns in satisfaction, functionality, and challenges. As detailed in the methodology, apps were categorized by function in Table 1. This underscores the market's emphasis on homework and math aids, with fewer dedicated tools in other areas, aligning with demand for targeted, immediate assistance in e-learning.

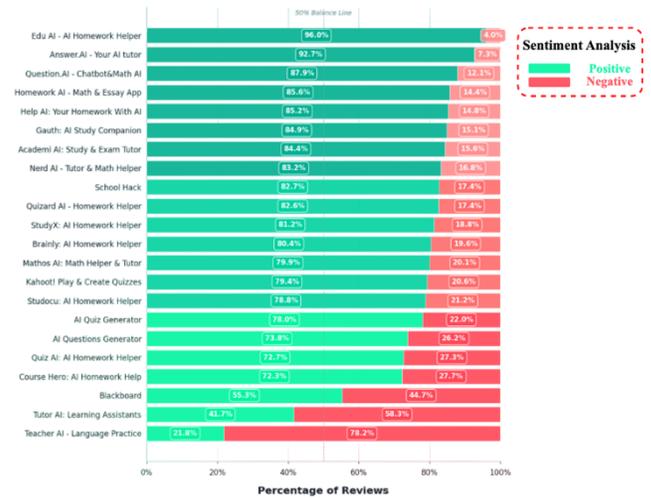

Figure 4. Positive and negative review percentages for AI educational apps

Figure 4 illustrates the distribution of positive and negative review percentages across the apps, ordered from highest to lowest positive sentiment. Overall reception is predominantly positive, with most apps exceeding 50% positive feedback. Top performers include Edu AI - Homework Helper (95.9% positive), Answer.AI - Your AI tutor (92.7%), and Question.AI - Chatbot&Math AI (87.9%), while lower ones are Blackboard (55.3%), Tutor AI: Learning Assistants (41.7%), and Teacher AI - Language Practice (21.8%). Homework helpers and multi-tool companions dominate high ranks, reflecting strong approval, whereas LMS and language apps exhibit greater negativity.

Key sentiments extracted from the reviews highlight positives centered on efficiency, accuracy, and personalization. For instance, apps like Edu AI and Answer.AI were frequently praised for fast responses, step-by-step explanations, and conceptual support in homework completion. Multi-tool and math-focused apps (e.g., Gauth, Nerd AI) received acclaim for quick problem-solving and reliable STEM assistance, while quiz generators like Kahoot! were noted for engaging gamification and customization. Conversely, negatives recurrently involved inaccuracies or incomplete answers (e.g., Quiz AI, Course Hero), aggressive paywalls and limited free access (e.g., Help AI, Quizard AI), intrusive ads (e.g., Brainly, Question.AI), technical instability such as crashes (e.g., Blackboard, Studocu), unreliable inputs like camera scanning (e.g., Nerd AI, Homework AI), and billing errors (e.g., Course Hero). Lower-performing apps like Tutor AI and Teacher AI faced criticism for outdated AI, poor speech recognition, and absent features like offline access.

Cross-app comparisons reveal category trends: Homework helpers averaged >80% positive sentiment for speed and versatility; quiz generators ~mid-70s, with content quality issues; language/LMS apps <60%, due to instability. Multifunctional apps (e.g., StudyX, Nerd AI) boosted retention but heightened monetization complaints, while specialized ones (e.g., Mathos AI) offered ad-free models but missed features like image support. Community apps (e.g., Brainly) succeeded moderately in collaboration yet were hindered, unlike gamified ones (e.g., Kahoot!) that excelled in engagement but faced connectivity issues. Math-focused apps (e.g., Homework AI, Mathos AI) were praised for equation handling but criticized for notation failures, highlighting AI domain lags. Overall, AI in e-teaching skews positive (~85% for homework vs. ~75% for specialized), though reliability and monetization negatives persist.

## V. DISCUSSION

The sentiment-driven evaluation of user reviews from top AI educational apps highlights generative AI's promise for transforming e-teaching, while revealing hurdles to adoption. Positive sentiments dominate, with homework helpers like Edu AI and multi-tool apps like Answer.AI excelling in personalized assistance, such as step-by-step explanations and quick problem-solving, aligning with student productivity needs [2]. This mirrors trends where students use AI for brainstorming and summarizing, democratizing resources for self-directed learning amid workloads [2]. Educators lag in adoption but gain from tools like lesson planning, reducing administrative burdens, and promoting inclusivity [3]. However, negatives like inaccuracies, paywalls, and instability expose gaps in capabilities, raising ethical issues such as over-reliance, plagiarism, and digital divides [6], [11]. Lower performers, including Teacher AI and Blackboard, demonstrate flaws in specialized features that erode trust and accessibility, especially in under-resourced areas [9].

These findings provide empirical, scalable insights from real-world reviews, revealing GenAI education apps' strong momentum in personalized learning—particularly homework/math tools (85% positive sentiment)—by bridging gaps in under-resourced areas, boosting creativity, efficiency, and engagement through gamification and community features [9]. However, multifunctional designs, while driving retention, amplify monetization barriers, ads, crashes, and equity issues, creating a double-edged sword that demands refinement for pedagogical alignment and ethical integration [11].

Looking ahead, the data points toward hybrid AI-human models as a pivotal evolution, combining app strengths like real-time assistance with teacher oversight to mitigate risks such as inaccuracies and dependency. For example, integrating high-rated features (e.g., Gauth's step-by-step explanations) with human-guided ethical checks could create collaborative systems where AI handles routine tasks, freeing educators for deeper interactions and fostering critical thinking [14], [15]. This hybrid approach addresses the cautious educator adoption by positioning AI as a supportive tool rather than a replacement, enhancing academic integrity and personalized pathways [5], [7].

Furthermore, untapped technologies like virtual reality (VR) and augmented reality (AR) offer opportunities for immersive learning, extending beyond current app limitations. VR/AR could simulate real-world scenarios—such as virtual labs for math solvers or interactive quizzes in Kahoot!—to bridge gaps in engagement and multimodal inputs, making education more accessible for diverse learners and filling voids in features like offline access or diagram support [16]. Our sentiment trends, with praises for engagement but criticisms of instability, suggest that embedding VR/AR in multifunctional apps could elevate user satisfaction, particularly in STEM and language domains where notation or speech recognition failures persist [8], [9].

To realize this vision, a practical roadmap is proposed: Developers should prioritize adaptive personalization (via feedback loops), expand collaborative features (currently underused beyond Kahoot!), mitigate ethical biases, integrate plagiarism detection and progress tracking, sustain gamification for engagement, and adopt transparent, equitable monetization (e.g., Mathos AI's ad-free model) instead of aggressive paywalls [11]. Policymakers should regulate for inclusivity by mandating expanded free tiers, multilingual support, accuracy standards, and strong data privacy protections to bridge digital divides [17, 18]. These measures can transform current descriptive tools into equitable, innovative e-teaching platforms accessible to all learners.

## VI. CONCLUSION

In conclusion, this sentiment-driven evaluation of user reviews from top AI educational apps on the Google Play Store highlights generative AI's transformative potential in e-teaching, with positive feedback emphasizing efficiencies in homework assistance, personalization, and engagement—especially in homework helpers (averaging 85% positives)—while revealing challenges like inaccuracies, monetization barriers, and technical instability that impede equitable adoption. Extending beyond analysis, it proposes hybrid AI-human models, VR/AR integrations for immersive learning, and a roadmap prioritizing adaptive features for developers and regulatory inclusivity for policymakers, offering empirical insights to bridge user-developer gaps and foster ethical refinements, mitigating risks such as dependency and digital divides. These findings advocate a shift to collaborative, innovative ecosystems empowering students and educators, guiding future research on longitudinal impacts and cross-platform integrations to democratize high-quality e-learning.